\documentclass[12pt,draftclsnofoot,onecolumn]{IEEEtran}
\usepackage{cite}
\usepackage[cmex10]{amsmath}
\usepackage{amssymb}
\usepackage{mathtools}
\usepackage{amsthm,thmtools}
\usepackage{url}
\usepackage[dvipsnames]{xcolor}
\usepackage{bm}
\usepackage{standalone}
\usepackage{enumitem}
\usepackage[yyyymmdd,hhmmss]{datetime}
\usepackage{tikz}
\usepackage{tikz-3dplot}
\usetikzlibrary{patterns}
\usetikzlibrary{shapes,arrows,positioning}
\usetikzlibrary{angles,quotes}
\usetikzlibrary{calc}
\usetikzlibrary{fit}
\usetikzlibrary{plotmarks}
\usetikzlibrary{decorations}
\tikzset{every picture/.style={>=stealth}}
\tikzset{three sided left/.style={
        draw=none,
        xshift=\pgflinewidth,
        append after command={
            [shorten <= -0.5\pgflinewidth]
            ([shift={(-1.5\pgflinewidth,-0.5\pgflinewidth)}]\tikzlastnode.north east) edge ([shift={( 0.5\pgflinewidth,-0.5\pgflinewidth)}]\tikzlastnode.north west) 
            ([shift={( 0.5\pgflinewidth,-0.5\pgflinewidth)}]\tikzlastnode.north west) edge ([shift={( 0.5\pgflinewidth,+0.5\pgflinewidth)}]\tikzlastnode.south west)            
            ([shift={( 0.5\pgflinewidth,+0.5\pgflinewidth)}]\tikzlastnode.south west) edge ([shift={(-1.0\pgflinewidth,+0.5\pgflinewidth)}]\tikzlastnode.south east)
        }}}
        
\tikzset{three sided right/.style={
        draw=none,
        xshift=-\pgflinewidth,
        append after command={
            [shorten <= -0.5\pgflinewidth]
            ([shift={( 1.5\pgflinewidth,-0.5\pgflinewidth)}]\tikzlastnode.north west) edge ([shift={(-0.5\pgflinewidth,-0.5\pgflinewidth)}]\tikzlastnode.north east) 
            ([shift={(-0.5\pgflinewidth,-0.5\pgflinewidth)}]\tikzlastnode.north east) edge ([shift={(-0.5\pgflinewidth,+0.5\pgflinewidth)}]\tikzlastnode.south east)            
            ([shift={(-0.5\pgflinewidth,+0.5\pgflinewidth)}]\tikzlastnode.south east) edge ([shift={( 1.0\pgflinewidth,+0.5\pgflinewidth)}]\tikzlastnode.south west)
        }}}

\usepackage{pgfplots}
\usepgfplotslibrary{polar}
\pgfplotsset{
  compat=newest, 
  width=0.60\columnwidth,    
  height=0.42\columnwidth,   
  plot coordinates/math parser=false,
  standard/.style={
    axis equal,
    axis line style=help lines,
    axis x line=center,
    axis y line=center,
    axis z line=center},
    grid style={dashed,gray},
    minor grid style={dotted,gray},
    major grid style={dotted,gray},
    ylabel absolute, ylabel style={yshift=-0.1cm},
    xlabel absolute, xlabel style={yshift=0.25cm}
}

\makeatletter
 \pgfdeclarepatternformonly[\tikz@pattern@color,\LineSpace,\LineWidth]{my horizontal lines}%
    {\pgfpointorigin}{\pgfqpoint{100pt}{1pt}}{\pgfqpoint{100pt}{\LineSpace}}%
    {
        \pgfsetcolor{\tikz@pattern@color}
        \pgfsetlinewidth{\LineWidth}
        \pgfpathmoveto{\pgfqpoint{0pt}{0.5pt}}
        \pgfpathlineto{\pgfqpoint{100pt}{0.5pt}}
        \pgfusepath{stroke}
    }
 
 \pgfdeclarepatternformonly[\tikz@pattern@color,\LineSpace,\LineWidth]{my vertical lines}%
    {\pgfpointorigin}{\pgfqpoint{1pt}{100pt}}{\pgfqpoint{\LineSpace}{100pt}}%
    {
        \pgfsetcolor{\tikz@pattern@color}
        \pgfsetlinewidth{\LineWidth}
        \pgfpathmoveto{\pgfqpoint{0.5pt}{0pt}}
        \pgfpathlineto{\pgfqpoint{0.5pt}{100pt}}
        \pgfusepath{stroke}
    } 
 
 \pgfdeclarepatternformonly[\tikz@pattern@color,\LineSpace,\LineWidth]{my grid}%
    {\pgfqpoint{-1pt}{-1pt}}{\pgfqpoint{\LineSpace}{\LineSpace}}
    {\pgfqpoint{\LineSpace}{\LineSpace}}%
    {
        \pgfsetcolor{\tikz@pattern@color}
        \pgfsetlinewidth{\LineWidth}
        \pgfpathmoveto{\pgfqpoint{0pt}{0pt}}
        \pgfpathlineto{\pgfqpoint{0pt}{\LineSpace + 0.1pt}}
        \pgfpathmoveto{\pgfqpoint{0pt}{0pt}}
        \pgfpathlineto{\pgfqpoint{\LineSpace + 0.1pt}{0pt}}
        \pgfusepath{stroke}
    }
 
 \pgfdeclarepatternformonly[\tikz@pattern@color,\LineSpace,\LineWidth]{my north east lines}
    {\pgfqpoint{-\LineWidth}{-\LineWidth}}{\pgfqpoint{\LineSpace}{\LineSpace}}
    {\pgfqpoint{\LineSpace}{\LineSpace}}%
    {
        \pgfsetcolor{\tikz@pattern@color}
        \pgfsetlinewidth{\LineWidth}
        \pgfpathmoveto{\pgfqpoint{-\LineWidth}{-\LineWidth}}
        \pgfpathlineto{\pgfqpoint{\LineSpace + 0.1pt}{\LineSpace + 0.1pt}}
        \pgfusepath{stroke}
    }
 
 \pgfdeclarepatternformonly[\tikz@pattern@color,\LineSpace,\LineWidth]{my north west lines}
    {\pgfqpoint{-\LineWidth}{-\LineWidth}}{\pgfqpoint{\LineSpace}{\LineSpace}}
    {\pgfqpoint{\LineSpace}{\LineSpace}}%
    {
        \pgfsetcolor{\tikz@pattern@color}
        \pgfsetlinewidth{\LineWidth}
        \pgfpathmoveto{\pgfqpoint{-\LineWidth}{\LineSpace}}
        \pgfpathlineto{\pgfqpoint{\LineSpace + 0.1pt}{-\LineWidth}}
        \pgfusepath{stroke}
    }
 
 \pgfdeclarepatternformonly[\tikz@pattern@color,\LineSpace,\LineWidth]{my crosshatch}%
    {\pgfqpoint{-1pt}{-1pt}}{\pgfqpoint{\LineSpace}{\LineSpace}}
    {\pgfqpoint{\LineSpace}{\LineSpace}}%
    {
        \pgfsetcolor{\tikz@pattern@color}
        \pgfsetlinewidth{\LineWidth}
        \pgfpathmoveto{\pgfqpoint{\LineSpace + 0.1pt}{0pt}}
        \pgfpathlineto{\pgfqpoint{0pt}{\LineSpace + 0.1pt}}
        \pgfpathmoveto{\pgfqpoint{0pt}{0pt}}
        \pgfpathlineto{\pgfqpoint{\LineSpace + 0.1pt}{\LineSpace + 0.1pt}}
        \pgfusepath{stroke}
    }
 
 \pgfdeclarepatternformonly[\tikz@pattern@color,\LineSpace,\PointSize]{my dots}%
    {\pgfqpoint{-\LineSpace*0.25}{-\LineSpace*0.25}}
    {\pgfqpoint{\LineSpace*0.25}{\LineSpace*0.25}}
    {\pgfqpoint{\LineSpace*0.75}{\LineSpace*0.75}}%
    {
        \pgfsetcolor{\tikz@pattern@color}
        \pgfpathcircle{\pgfqpoint{0pt}{0pt}}{\PointSize}
        \pgfusepath{fill}
    }
    
\def\calcLength(#1,#2)#3{%
\pgfpointdiff{\pgfpointanchor{#1}{center}}%
             {\pgfpointanchor{#2}{center}}%
\pgf@xa=\pgf@x%
\pgf@ya=\pgf@y%
\FPeval\@temp@a{\pgfmath@tonumber{\pgf@xa}}%
\FPeval\@temp@b{\pgfmath@tonumber{\pgf@ya}}%
\FPeval\@temp@sum{(\@temp@a*\@temp@a+\@temp@b*\@temp@b)}%
\FProot{\FPMathLen}{\@temp@sum}{2}%
\FPround\FPMathLen\FPMathLen5\relax
\global\expandafter\edef\csname #3\endcsname{\FPMathLen}
}    
    
\makeatother
 
\newdimen\LineSpace
\newdimen\PointSize
\newdimen\LineWidth
\tikzset{
    line space/.code={\LineSpace=#1},
    line space=3pt
}
\tikzset{
    point size/.code={\PointSize=#1},
    point size=.5pt
}
\tikzset{
    pattern line width/.code={\LineWidth=#1},
    pattern line width=.4pt
}

\tikzset{antenna/.style={insert path={-- coordinate (ant#1) ++(0,0.25) -- +(135:0.25) + (0,0) -- +(45:0.25)}}}

\pgfdeclaredecoration{discontinuity}{start}{
  \state{start}[width=0.5\pgfdecoratedinputsegmentremainingdistance-0.5\pgfdecorationsegmentlength,next state=first wave]
  {}
  \state{first wave}[width=\pgfdecorationsegmentlength, next state=second wave]
  {
    \pgfpathlineto{\pgfpointorigin}
    \pgfpathmoveto{\pgfqpoint{0pt}{\pgfdecorationsegmentamplitude}}
    \pgfpathcurveto
        {\pgfpoint{-0.25*\pgfmetadecorationsegmentlength}{0.75\pgfdecorationsegmentamplitude}}
        {\pgfpoint{-0.25*\pgfmetadecorationsegmentlength}{0.25\pgfdecorationsegmentamplitude}}
        {\pgfpoint{0pt}{0pt}}
    \pgfpathcurveto
        {\pgfpoint{0.25*\pgfmetadecorationsegmentlength}{-0.25\pgfdecorationsegmentamplitude}}
        {\pgfpoint{0.25*\pgfmetadecorationsegmentlength}{-0.75\pgfdecorationsegmentamplitude}}
        {\pgfpoint{0pt}{-\pgfdecorationsegmentamplitude}}
}
\state{second wave}[width=0pt, next state=do nothing]
  {
    \pgfpathmoveto{\pgfqpoint{0pt}{\pgfdecorationsegmentamplitude}}
    \pgfpathcurveto
        {\pgfpoint{-0.25*\pgfmetadecorationsegmentlength}{0.75\pgfdecorationsegmentamplitude}}
        {\pgfpoint{-0.25*\pgfmetadecorationsegmentlength}{0.25\pgfdecorationsegmentamplitude}}
        {\pgfpoint{0pt}{0pt}}
    \pgfpathcurveto
        {\pgfpoint{0.25*\pgfmetadecorationsegmentlength}{-0.25\pgfdecorationsegmentamplitude}}
        {\pgfpoint{0.25*\pgfmetadecorationsegmentlength}{-0.75\pgfdecorationsegmentamplitude}}
        {\pgfpoint{0pt}{-\pgfdecorationsegmentamplitude}}
    \pgfpathmoveto{\pgfpointorigin}
}
  \state{do nothing}[width=\pgfdecorationsegmentlength,next state=do nothing]{
    \pgfpathlineto{\pgfpointdecoratedinputsegmentlast}
  }
  \state{final}
  {
    \pgfpathlineto{\pgfpointdecoratedpathlast}
  }
}

\newcommand{\pgfextractangle}[3]{%
    \pgfmathanglebetweenpoints{\pgfpointanchor{#2}{center}}
                              {\pgfpointanchor{#3}{center}}
    \global\let#1\pgfmathresult  
}

\usepackage{silence}
\usepackage{glossaries}
\newacronym{OFDM}{OFDM}{orthogonal frequency division multiplexing}
\newacronym{QPSK}{QPSK}{QPSK}
\newacronym{BPSK}{BPSK}{BPSK}
\newacronym{MAC}{MAC}{medium access control}
\newacronym{PHY}{PHY}{physical}
\newacronym{CP}{CP}{cyclic prefix}
\newacronym{CRC}{CRC}{cyclic redundancy check}
\newacronym{LS}{LS}{least-squares}
\newacronym{FER}{FER}{frame error rate}
\newacronym{MCS}{MCS}{modulation and coding scheme}
\newacronym{LLR}{LLR}{log-likelihood ratio}
\newacronym{FPGA}{FPGA}{field programmable gate array}
\newacronym{PDP}{PDP}{power delay profile}
\newacronym{SNR}{SNR}{signal-to-noise ratio}
\newacronym{CSI}{CSI}{channel state information}
\newacronym{ADC}{ADC}{analog to digital converter}
\newacronym{DAC}{DAC}{digital to analog converter}
\newacronym{LMMSE}{LMMSE}{linear minimum mean squared error}
\newacronym{LO}{LO}{local oscillator}
\newacronym{LNA}{LNA}{low noise amplifier}
\newacronym{PA}{PA}{power amplifier}
\newacronym{LPF}{LPF}{low pass filter}
\newacronym{SISO}{SISO}{single input single output}
\newacronym{MIMO}{MIMO}{multiple input multiple output}
\newacronym{MSE}{MSE}{mean squared error}
\newacronym{AOA}{AOA}{angle of arrival}
\newacronym{TX}{TX}{transmitter}
\newacronym{RX}{RX}{receiver}
\newacronym{PEP}{PEP}{packet error probability}
\newacronym{EGC}{EGC}{equal gain combining}
\newacronym{AWGN}{AWGN}{additive white Gaussian noise}
\newacronym{CAM}{CAM}{cooperative awareness messages}
\newacronym{MRC}{MRC}{maximal ratio combining}
\newacronym{SC}{SC}{selection combining}
\newacronym{BEP}{BEP}{burst error probability}
\newacronym{ACN}{ACN}{analog combining network}
\newacronym{BSM}{BSM}{basic safety message}

\theoremstyle{remark}
\declaretheorem[name=Example,qed={\lower-0.3ex\hbox{$\triangle$}}]{example}

\declaretheoremstyle[
spaceabove=6pt, spacebelow=6pt,
headfont=\normalfont\bfseries,
notefont=\mdseries, notebraces={(}{)},
bodyfont=\normalfont,
postheadspace=1em,
numberwithin=section,
qed=$\blacktriangleleft$
]{lemmastyle}

\declaretheorem{lemma}
\declaretheorem{theorem}

\newcommand{\sgn}{\mathop{\mathrm{sgn}}\nolimits}
\newcommand{\E}{\mathop{\mathbb{E}}\nolimits}
\newcommand{\lr}[1]{\ensuremath{\left({#1}\right)}}
\newcommand{\lrb}[1]{\left \lbrace {#1} \right \rbrace}
\newcommand{\set}[1]{\left \lbrace {#1} \right \rbrace}
\newcommand{\lrh}[1]{\left [ {#1} \right ]}

\newcommand{\qfunc}[1]{Q \left( #1 \right) }
\newcommand{\mymod}[2]{\text{mod}\left( #1,#2 \right)}
\newcommand{\rpm}{\sbox0{$1$}\sbox2{$\scriptstyle\pm$} \raise\dimexpr(\ht0-\ht2)/2\relax\box2 }

\newcommand{\figref}[1]{Fig.~\ref{#1}}

\newcommand{\abs}[1]{\left\lvert#1\right\rvert}

\newcommand\numberthis{\addtocounter{equation}{1}\tag{\theequation}}

\DeclarePairedDelimiter{\ceil}{\lceil}{\rceil}  

\newcommand{\valf}{\bm{\alpha}}
\newcommand{\vpsi}{\bm{\psi}}
\newcommand{\vbet}{\bm{\beta}}

\begin{document}      
\title{Robust Connectivity with Multiple Nonisotropic Antennas for Vehicular Communications}
    	
\author{\IEEEauthorblockN{
    Keerthi Kumar Nagalapur,
    Erik G. Str\"om,
    Fredrik Br\"annstr\"om,
    Jan Carlsson,
    and Kristian Karlsson}
\thanks{
Keerthi Kumar Nagalapur, Fredrik Br\"annstr\"om, and Erik G. Str\"om are with the Division of Communication Systems, Department of Signals and Systems, Chalmers University of Technology, SE-412 96, Gothenburg, Sweden, E-mail: \emph{\{keerthi, fredrik.brannstrom, erik.strom\}@chalmers.se}. 
Jan Carlsson is with Provinn AB and Department of Signals and Systems, Chalmers University of Technology, Gothenburg, Sweden, E-mail: \emph{jan.carlsson@provinn.se}.
Kristian Karlsson is with the Department of Electronics, SP Technical Research Institute of Sweden, Bor{\aa}s, Sweden, E-mail: \emph{kristian.karlsson@sp.se}.}
\thanks{The research was partially funded by Swedish Governmental Agency for Innovation Systems (VINNOVA) within the VINN Excellence Center Chase project \emph{Antenna Systems for V2X Communication.}}      
}
\maketitle

\begin{abstract}
For critical services, such as traffic safety and traffic efficiency, it is advisable to design systems with robustness as the main criteria, possibly at the price of reduced peak performance and efficiency.
Ensuring robust communications in case of embedded or hidden antennas is a challenging task due to nonisotropic radiation patterns of these antennas.
The challenges due to the nonisotropic radiation patterns can be overcome with the use of multiple antennas.
In this paper, we describe a simple, low-cost method for combining the output of multiple nonisotropic antennas to guarantee robustness, i.e., support reliable communications in worst-case scenarios.
The combining method is designed to minimize the burst error probability, i.e., the probability of consecutive decoding errors of status messages arriving periodically at a receiver from an arbitrary angle of arrival.
The proposed method does not require the knowledge of instantaneous signal-to-noise ratios or the complex-valued channel gains at the antenna outputs.
The proposed method is applied to measured and theoretical antenna radiation patterns, and it is shown that the method supports robust communications from an arbitrary angle of arrival.
\end{abstract}

\vspace{2\baselineskip}

\section{Introduction}
\label{sec:intro}
Vehicular traffic safety and traffic efficiency applications demand reliable and robust communication between vehicles.
These applications are enabled by vehicles that transmit periodic status messages, referred to as \glspl{CAM} in Europe and \glspl{BSM} in the US~\cite{ETSI_CAM,SAEJ2735}, containing current position, speed, heading, etc.
A shark fin antenna module located on top of a vehicle's roof is the standard method for housing the antennas used for vehicular communications today.
Conformal/hidden antennas are being considered instead of the shark fin modules for the reasons of safety of the antennas, exterior appearance of the vehicle, and aerodynamics.
Radiation patterns of the hidden antennas are typically nonisotropic due to the vehicle components that closely surround them.
The resulting nonisotropic patterns might have very low power gains in certain angles or in the worst-case scenario even nulls.
If the signal from a \gls{TX} arrives at a \gls{RX} in a very narrow sector and the \gls{AOA} of the signal coincides with one of the angles of the receiving antenna having a low gain, it might not be possible to decode the transmitted packet successfully due to the resulting low \gls{SNR}.
Since the position of a vehicle varies slowly over the time duration of a few consecutive packets, we can expect the \gls{AOA} of the signal from the vehicle to remain approximately the same over this duration.
As a result, there is a risk of a sequence of consecutive packets arriving at an \gls{AOA} coinciding with one of the angles corresponding to low gains in the nonisotropic antenna pattern.

The problems due to nonisotropic antenna patterns can be remedied by using multiple antennas with contrasting radiation patterns.
Combining the outputs of the multiple antennas is a well studied topic and methods such as \gls{SC}, \gls{EGC}, and \gls{MRC} have been investigated thoroughly~\cite{Goldsmith2005WL}.
These methods either require the knowledge of the instantaneous channel amplitude and phase, or the SNR of the output signal on each antenna branch.
Schemes that do not require the aforementioned information for combining have also been studied.
A scheme called random beamforming has been explored in~\cite{Yang2012}, where the antenna pattern is randomized over several time-frequency blocks to achieve omnidirectional coverage on average.

Typically, the combining methods described above require an \gls{ADC} on each of the antenna branches and a multiport \gls{RX} to combine the signals digitally.
The multiport \gls{RX} uses either the \gls{SNR} or complex channel gain of the signal on each of the ports to combine the signals.
An alternative to this approach is to use an \gls{ACN} consisting of analog phase shifters, variable gain amplifiers, and combiners to obtain a single combined signal that requires only a single \gls{ADC} together with a single port \gls{RX}~\cite{VijayVeen2010,Gholam2011}.
When the antennas and the \gls{RX} are co-located, it is convenient to use a closed loop system where the information from the \gls{RX} is used to control the analog combining network. 
When the \gls{ACN} does not receive any feedback from the \gls{RX}, it can be designed to satisfy some performance criterion. 
Such an \gls{ACN} can be designed as an integrated part of the antenna system independent of the \gls{RX}.

In this work, we investigate an \gls{ACN} that does not receive any feedback from the \gls{RX}.
Furthermore, the \gls{ACN} is designed to operate without the knowledge of the instantaneous complex channel gain and/or \gls{SNR} of the branches to keep the implementation complexity to a minimum.
The \gls{ACN} aims to provide robust connectivity by exploiting the periodic nature of the \glspl{CAM}.
Although a \gls{TX} periodically broadcasts \glspl{CAM}, it might not be strictly required that every message is successfully decoded for applications to work as intended,
since the \glspl{CAM} contain information of the physical quantities that vary slowly over the time duration of few packets. 
However, losing a number of consecutive packets will have serious implications on the functioning of the applications.
A simple model to capture this behavior is to declare an application outage if $K>1$ consecutive packets are not successfully decoded. 
The communication system should then be designed to minimize the \gls{BEP}, i.e., the probability that a burst of $K$ consecutive packet errors occurs.
Therefore, we design our antenna combining method to minimize the \gls{BEP}.
Minimizing the \gls{BEP} is equivalent to minimizing the probability of zero successful packets decoded in $\tau=K T$, where $T$ is the time between two consecutive \glspl{CAM}.
The duration $\tau$ can be viewed as the maximum duration between two successful packet receptions an application can tolerate before ceasing to function normally.

\section{System Model}
\begin{figure}
\centering
\includegraphics[scale=1]{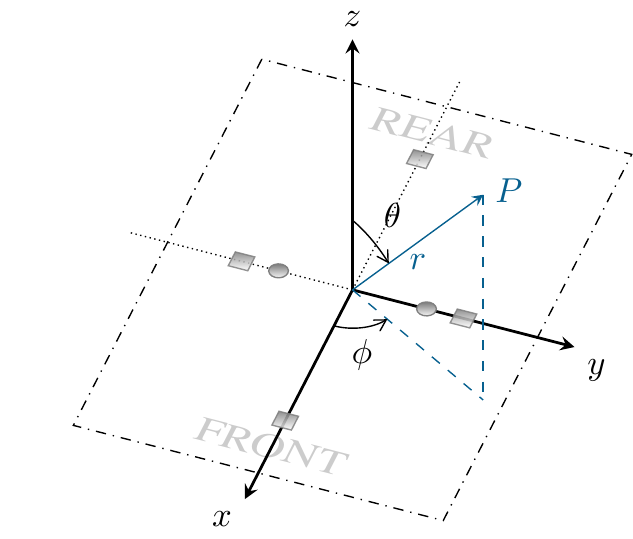}
\caption{Antenna coordinate system. The roof of the vehicle is in the $xy$ plane.}
\label{fig:coord}
\end{figure}

Consider $L \geq 2$ antennas located on a vehicle.
All the antennas are assumed to be at the same height from the ground and in the $xy$ plane as shown in \figref{fig:coord}.
The angles $\phi$ and $\theta$ are the azimuth and elevation angles, respectively.
Orientation of the vehicle with respect to the coordinate system is also shown in the figure.
Let $g_l(\phi)$ be the far-field function/response of the antenna $l \in \set{0,1,\dots,L-1}$ in the azimuth plane.
The far-field function is normalized such that $\abs{g_l(\phi)}^2$ represents the relative directive gain of the $l$th antenna with respect to an isotropic antenna.
The far-field function in the elevation plane has been omitted since we restrict the arrival of the waves to the azimuth plane, i.e., $\theta=\pi/2$.
For simplicity, we assume that the antennas are vertically polarized in the azimuth plane and that the incident electrical field is also vertically polarized.
Two examples of antenna placement are also shown in the figure (circles and squares). 

\label{sec:sysmodel}

The complex-valued channel gain at the $l$th antenna is given by~\cite[Eqn. 8]{Karedal2009}
\begin{align}
\label{eq:sysmod}
h_l(t) &= \sum_{n=1}^{N} \tilde{a}_n(t) g_l(\phi_n) e^{-\jmath \tilde{\Omega}_{n,l}(t) },
\end{align}
where $\tilde{a}_n(t)$ is the complex-valued gain of the $n$th multipath component having an \gls{AOA}, $\phi_n$; 
$e^{-\jmath \tilde{\Omega}_{n,l}(t)}$ is the distance-induced phase shift of the $n$th component at the $l$th antenna such that $\tilde{\Omega}_{n,l}(t)= (2\pi/\lambda) d_{n,l}(t)$, where $d_{n,l}(t)$ is the
time-varying propagation distance of the $n$th component at the $l$th antenna and $\lambda$ is the wavelength of the carrier signal.

Considering the $l=0$ antenna as the reference, the channel gain at the $l$th element can be written as
\begin{align}
h_l(t) &=  \sum_{n=1}^{N} a_n(t) g_l(\phi_n) e^{-\jmath \Omega_{n,l}(t)},
\end{align}
where $a_n(t) = \tilde{a}_n(t) e^{-\jmath \tilde{\Omega}_{n,0}(t)}$ and $\Omega_{n,l}(t) = \tilde{\Omega}_{n,l}(t)-\tilde{\Omega}_{n,0}(t)$ is the relative phase difference experienced by the $n$th component at the $l$th antenna with respect to the reference antenna.

\begin{figure}
\centering
\includegraphics[scale=1]{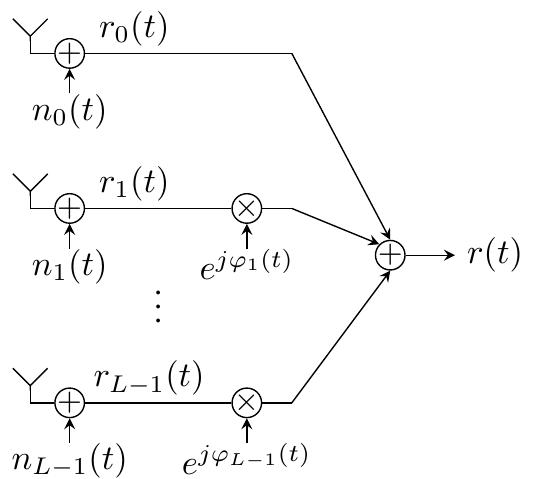}
\caption{The analog combining network with $L$ antennas.}
\label{fig:antsys}
\end{figure}

The signal at the output of the $l$th antenna is given by
\begin{align}
r_l(t) &= s(t) h_l(t) + n_l(t),
\end{align}
where $s(t)$ is the transmitted signal and $n_l(t)$ is independent complex \gls{AWGN} at the $l$th antenna having an average power $\E\{\abs{n_l(t)}^2\}=P_{\rm{n}}, \forall l$ with respect to the bandwidth of the signal $s(t)$.
We restrict the \gls{ACN} to consist of analog phase shifters and an adder as seen in \figref{fig:antsys}. 
The outputs of the $L-1$ antennas are phase rotated and added to the output of the reference antenna. 
The output of the combiner $r(t)$ is given by
\begin{align}
r(t) &= \mathop{\sum}_{l=0}^{L-1} r_l(t) e^{\jmath \varphi_l(t)},
\end{align}
where $\varphi_l(t)$ is the time-varying phase shift applied to the $l$th antenna output and $\varphi_0(t)=0$ since the output of the reference antenna is not phase rotated. 
Since the \gls{ACN} does not use any information from the \gls{RX} and the signal \gls{SNR} is not measured, $\varphi_l(t)$ as a function of time has to be predetermined according to some performance criterion.
To simplify the design of $\varphi_l(t)$, which is a continuous function of time, we propose to model it as a linear function of time,
i.e., $\varphi_l(t)= \alpha_l t+\beta_l$, where $\alpha_l$ is the slope and $\beta_l$ is the phase offset of the $l$th phase shifter.
Since the output of the reference antenna is not phase rotated, it follows that $\alpha_0=\beta_0=0$.
The output of the combiner is given by
\begin{align*}
r(t)&=\mathop{\sum}_{l=0}^{L-1} r_l(t) e^{\jmath (\alpha_l t + \beta_l)} \\
    &= s(t) \!\mathop{\sum}_{l=0}^{L-1} h_l(t) e^{\jmath (\alpha_l t + \beta_l)} + \!\mathop{\sum}_{l=0}^{L-1}\!n_l(t) e^{\jmath (\alpha_l t + \beta_l)} \\
    &= s(t) h(\valf, \vbet, t) + \mathop{\sum}_{l=0}^{L-1} \tilde{n}_l(t), \numberthis \label{eq:recsig}
\end{align*}
where $h(\valf, \vbet, t)$ is the effective time-varying channel, $\valf = \lrh{\alpha_1,\alpha_2,\ldots,\alpha_{L-1}}^{\mathsf{T}}$, and $\vbet=\lrh{\beta_1,\beta_2,\ldots,\beta_{L-1}}^{\mathsf{T}}$; $\tilde{n}_l(t)$ is also complex additive white Gaussian noise with average power $P_{\rm{n}}$ since $n_l(t)$ is circularly symmetric.

The received signal in~\eqref{eq:recsig}, consequently the \gls{SNR}, is a function of the complex channel coefficients $h_l(t)$ and the time-varying phase shifts $\varphi_l(t)$. 
The time-varying phase shifts have to be determined to minimize the \gls{BEP} of the \glspl{CAM} based on the characteristics of $h_l(t)$.
The \glspl{CAM} are broadcast periodically with a period of $T \, \rm{s}$.
Using IEEE 802.11p with a throughput of $6$ Mb/s as the reference physical layer~\cite{IEEE802112012}, the duration of a packet $T_{\rm{P}}$ is in the order of $0.5$ to $2$ ms corresponding to packet sizes of approximately $400$ to $1500$ bytes.
This duration is very small in comparison to $T$ which is in the order of $0.1$ to $1$ s~\cite[Table 1]{ETSI_CAM}.
The duration of a packet and the period of the \glspl{CAM} have to be considered while determining the time-varying phase shifts.
Having noted the nature of the \glspl{CAM}, we consider two contrasting vehicular channel models and discuss the implication of the proposed combining scheme on the \glspl{CAM} in the considered models.

\vspace{0.5\baselineskip}
\noindent \textbf{Scenario 1}, \emph{a single line-of-sight path between the \gls{TX} and the \gls{RX} with an \gls{AOA} $\phi$:} 
this is a reasonable model for highway environments, which typically have few scatterers and therefore relatively few multipath components contributing to the received power.
In this scenario, the channel at the $l$th antenna is given by 
\begin{align}
\label{eq:LoS}
h_l(t) = a(t) g_l(\phi) e^{-\jmath \Omega_{l}(t)},
\end{align}
where the indexing in $n$ is omitted due to a single component. 
Furthermore, when multiple paths arrive at the \gls{RX} with a very narrow angular spread centered around angle $\phi$, the channel can be approximated as a single line-of-sight path.
This scenario occurs in highway environments when the \gls{TX} is surrounded by local scatterers and the \gls{RX} is at a large distance from the \gls{TX}. 
When the angular spread is narrow such that $\abs{d_{n,l}-d_{l}} \approx 0$ where $d_l = (1/N)\sum_n d_{n,l}$ and $g_l(\phi_n) \approx g_l(\phi)\, \forall n$, the complex gain at the $l$th antenna can be approximated by the right hand side of \eqref{eq:LoS}, where $a(t) = \sum_{n=1}^{N} a_n(t)$, $\Omega_{l}(t) = \tilde{\Omega}_{l}(t) - \tilde{\Omega}_{0}(t)$ and $\tilde{\Omega}_l(t) = (2\pi/\lambda)d_l$.
The approximation is invalid for large angular spreads.

As seen from~\eqref{eq:LoS}, the equivalent channel at the $l$th antenna suffers from high attenuation when the antenna has low gain at the \gls{AOA} $\phi$.
Since the \gls{AOA} remains approximately constant over the duration of $K$ consecutive packets, the risk of losing all the $K$ packets is high.
It is possible to alleviate the problem by combining the output of multiple antennas with contrasting far-field functions.
Therefore, we focus on designing the \gls{ACN} to reduce the \gls{BEP} in this scenario.

The output of the combiner $r(t)$ is given by
\begin{align*}
r(t)&= s(t)a(t)\!\mathop{\sum}_{l=0}^{L-1}\!g_l(\phi) e^{-\jmath\lr{  \Omega_{l}(t) -\alpha_l t-\beta_l}  } 
      + \mathop{\sum}_{l=0}^{L-1} \tilde{n}_l(t) \\  
	&= s(t)a(t) g(\phi,\valf,\vbet,t) + \mathop{\sum}_{l=0}^{L-1} \tilde{n}_l(t), \numberthis
\end{align*}
where $g(\phi,\valf,\vbet,t)$ is the effective time-varying antenna far-field function.

The rates of phase shift $\alpha_l$ have to be chosen such that the time variation of $g(\phi,\valf,\vbet,t)$ over the duration of a \gls{CAM} packet is negligible and the variation between two consecutive packets arriving from a \gls{TX} with the period $T$ is significant enough.
When the phase shift over a packet is negligible, $g(\phi,\valf,\vbet,t)$ remains approximately constant over the duration of a packet.
Therefore, the effective far-field function during the $k$th packet can be approximated to be $g(\phi,\valf,\vbet,kT)$.
Consequently, the average \gls{SNR} of the $k$th packet is given by
\begin{align}
\label{eqn:effrad}
\bar{\gamma}(\phi,\valf,\vbet,k)\!&=\!\frac{ \E\lrb{|a(t) s(t)|^2}  |g(\phi,\valf,\vbet,kT)|^2 } { \E\lrb{\abs{\mathop{\sum}_{l=0}^{L-1} \tilde{n}_l(t)}^2}}.
\end{align}

Over the duration $\tau = KT$, when $K$ is in the order of $5$ to $10$, the path-loss between the \gls{TX} and \gls{RX}, and the \gls{AOA} $\phi$ approximately remain the same.
Therefore, for the $K$ consecutive packets under consideration the average received power can be assumed to be constant and given as $\E\lrb{|a(t) s(t)|^2} = P_{\rm{r}}$. 
The average \gls{SNR} of the $k$th packet is then given by
\begin{align}
\label{eq:snrpkt}
\bar{\gamma}(\phi,\valf,\vbet,k) &= \frac{P_{\rm{r}}}{LP_{\rm{n}}} |g(\phi,\valf,\vbet,kT)|^2.
\end{align}

The assumption that the \gls{AOA} $\phi$ remains approximately constant may be invalid when the distance between the \gls{TX} and \gls{RX} is small,
and the \gls{TX} and \gls{RX} are moving with high relative velocities.
However, in this scenario the received power is significantly higher due the smaller \gls{TX}-\gls{RX} separation and is therefore not a limiting scenario for the applicability of the proposed scheme (which aims to improve reception in the low-\gls{SNR} regime).

\vspace{0.5\baselineskip}
\noindent \textbf{Scenario 2}, \emph{large number of multipath components arriving isotropically, i.e., $N \gg 1$ and $\phi_n$ is uniformly distributed in the interval $[0,2\pi)$}: 
this scenario commonly occurs in urban environments and/or when the \gls{TX} and the \gls{RX} are surrounded by many vehicles acting as scatterers.
In this scenario, the channel gains $h_l(t)$ can be assumed to be uncorrelated complex Gaussian processes with zero mean when all of the following assumptions are satisfied:
(i) the separation between the antennas is larger than $\lambda$,
(ii) a dominant component is absent, and
(iii) the antennas are assumed to be isotropic~\cite[Sec. 5.4]{Molisch2010WL}.
The requirement of isotropic antennas in assumption (iii) can be relaxed when the antennas have a broad beamwidth or when the main lobes of the antennas are oriented in different directions. 

The signal at the output of the combiner is given by
\begin{align*}
r(t) &= s(t) \mathop{\sum}_{l=0}^{L-1} h_l(t) e^{\jmath (\alpha_l t + \beta_l)} + \mathop{\sum}_{l=0}^{L-1} n_l(t) e^{\jmath (\alpha_l t + \beta_l)}   \\
	 &= s(t) \mathop{\sum}_{l=0}^{L-1} \tilde{h}_l(t) + \mathop{\sum}_{l=0}^{L-1} \tilde{n}_l(t) \\
	 &= s(t) h(t) + \mathop{\sum}_{l=0}^{L-1} \tilde{n}_l(t), \numberthis
\end{align*}
where $\tilde{h}_l(t)$ is also complex Gaussian process due to the circular symmetry and $h(t)=\mathop{\sum}_{l=0}^{L-1} \tilde{h}_l(t)$ is the equivalent channel gain. 
Assuming that the path-loss and the large-scale fading between the \gls{TX} and the \gls{RX} are approximately constant over the duration of $K$ packets, the average \gls{SNR} of the $K$ packets is given by
\begin{align}
\bar{\gamma} &= \frac{ \E\lrb{ \abs{s(t) h(t)}^2 }} {\E\lrb{\abs{\mathop{\sum}_{l=0}^{L-1} \tilde{n}_l(t)}^2}} = \frac{\mathop{\sum}_{l=0}^{L-1} P_{\mathrm{r},l} }{L P_{\rm{n}}},			 
\end{align}
where $P_{\mathrm{r},l} = \E\lrb{ \abs{s(t) h_l(t)}^2 }$ and the \gls{SNR} is exponentially distributed with mean $\bar{\gamma}$. 
Suppose $\bar{\gamma}_{\rm{ISO}}$ is the mean \gls{SNR} of an isotropic antenna, when $\bar{\gamma} \geq \bar{\gamma}_{\rm{ISO}}$ the performance of the \gls{ACN} is better or equivalent to the performance of the isotropic antenna. 
Under the assumption of satisfying the above mentioned condition, the \gls{ACN} does not degrade the performance in the isotropic arrival scenario with respect to the single isotropic antenna.

\section{Burst Error Probability}
\label{sec:burstprob}
In this section, we formulate the problem of designing the \gls{ACN} to minimize the \gls{BEP} in the first scenario described in Section~\ref{sec:sysmodel}. 
The \gls{PEP} of the $k$th packet is a function of the average SNR and is denoted by $P_{\mathrm{e}}(\bar{\gamma}(\phi,\valf,\vbet,k))$.
The function $P_{\mathrm{e}}(\cdot)$ depends on the modulation and coding scheme used, the length of the packet, and the characteristics of the channel.
As mentioned earlier, we intend to minimize the probability of having a burst of $K$ consecutive packet errors denoted by $P_{\rm{B}}(\phi,\valf,\vbet,K)$.
Assuming that the packet errors are independent, the \gls{BEP} is given by
\begin{equation}
P_{\rm{B}}(\phi,\valf,\vbet,K) = \mathop{\prod}_{k=0}^{K-1} P_{\mathrm{e}}(\bar{\gamma}(\phi,\valf,\vbet,k)).
\end{equation}
Since we are interested in determining the optimum $\valf$ that minimizes the \gls{BEP} for the worst-case \gls{AOA} $\phi \in [0,2\pi)$, we formulate the following problem.
\begin{align}
\label{eq:allpep}
\valf^* &= \mathop{\arg \inf }_{\alpha_l \in \mathbb{R}} \mathop{\sup}_{\phi,\beta_l \in [0,2\pi)} P_{\rm{B}}(\phi,\valf,\vbet,K).
\end{align}
Note that we maximize the \gls{BEP} with respect to $\vbet$ in addition to $\phi$ to include the effect of the worst-case initial offset of $\varphi_l(t)$.

We now have a framework to find $\valf^*$ that minimizes the \gls{BEP} for arbitrary far-field functions of the antennas and \gls{PEP} functions when the signal arrives at the \gls{RX} as a single component or when the spread of the \gls{AOA} is very small.
It might not be possible to solve the optimization problem in~\eqref{eq:allpep} analytically for any given far-field function and \gls{PEP} function, in which case numerical optimization can be used.

As a special case of \gls{PEP} function, we consider an exponential \gls{PEP} function of the form $P_{\mathrm{e}}(\bar{\gamma})=a\exp(-b\bar{\gamma})$, where $a, b > 0$ are constants.
The \gls{BEP} in the case of the exponential \gls{PEP} function is given by
\begin{align*}
P_{\rm{B}}(\phi,\valf,\vbet,K) &= \mathop{\prod}_{k=0}^{K-1} a e^{ -b \bar{\gamma}(\phi,\valf,\vbet,k)}, \numberthis \\
\mathop{\ln}\left( P_{\rm{B}}(\phi,\valf,\vbet,K) \right) &= K \ln(a) - b \mathop{\sum}_{k=0}^{K-1} \bar{\gamma}(\phi,\valf,\vbet,k).
\end{align*}
The optimization problem in~\eqref{eq:allpep} can then be written as
\begin{align*}
\valf^*     &= \mathop{\arg \inf}_{\alpha_l \in \mathbb{R}} \mathop{\sup}_{\phi,\beta_l \in [0,2\pi)} \mathop{\ln}\left( P_{\rm{B}}(\phi,\valf,\vbet,K) \right) \\
            &= \mathop{\arg \sup}_{\alpha_l \in \mathbb{R}} \mathop{\inf}_{\phi,\beta_l \in [0,2\pi)} \mathop{\sum}_{k=0}^{K-1} \bar{\gamma}(\phi,\valf,\vbet,k) \numberthis \label{eq:sumofsnrs} \\
            &= \mathop{\arg \sup}_{\alpha_l \in \mathbb{R}} \mathop{\inf}_{\phi,\beta_l \in [0,2\pi)} \mathop{\sum}_{k=0}^{K-1} \frac{P_{\rm{r}}}{LP_{\rm{n}}} |g(\phi,\valf,\vbet,kT)|^2 \\
            &= \mathop{\arg \sup}_{\alpha_l \in \mathbb{R}} \mathop{\inf}_{\substack{ \phi,\beta_l \\ \in [0,2\pi) }} \mathop{\sum}_{k=0}^{K-1} \abs{ \sum_{l=0}^{L-1} g_l(\phi)e^{-\jmath \lr{ \Omega_{l}(kT) - \alpha_l kT - \beta_l }}  }^2.            
\end{align*} 
When the distance between the \gls{TX} and the \gls{RX} is large, and the separation between the antennas is not large, the relative phase difference $\Omega_{l}(kT)$ does not vary significantly over the duration of the $K$ packets we are considering.
As a consequence, we use the approximation $\Omega_{l}(kT) \approx \Omega_l, \; \forall k=\{0,1,\dots,K-1\}$.

Let $\psi_l = \mymod{ \Omega_{l} - \beta_l - \angle{g_l(\phi)} }{2\pi}$, where $\mymod{u}{v}$ is the remainder after dividing $u$ by $v$. Now, $\beta_l \in [0,2\pi)$ implies that $\psi_l \in [0,2\pi)$ and the optimization problem can be written as
\begin{flalign*}
& \valf^*=\mathop{\arg \sup}_{\alpha_l \in \mathbb{R}} \mathop{\inf}_{\phi,\psi_l \in [0,2\pi)} \mathop{\sum}_{k=0}^{K-1} \abs{ \sum_{l=0}^{L-1} \abs{g_l(\phi)} e^{-\jmath \lr{ \psi_l - \alpha_l kT}}  }^2 \\
		  &=\mathop{\arg \sup}_{\alpha_l \in \mathbb{R}} \mathop{\inf}_{\phi,\psi_l \in [0,2\pi)} \mathop{\sum}_{k=0}^{K-1} 
            \Bigg\{\!\lr{\sum_{l=0}^{L-1} \abs{g_l(\phi)} \cos\lr{\psi_l - \alpha_l kT}}^2 
            + \lr{\sum_{l=0}^{L-1} \abs{g_l(\phi)} \sin\lr{\psi_l - \alpha_l kT }}^2\!\Bigg\} \\          	
		  &= \mathop{\arg \sup}_{\alpha_l \in \mathbb{R}} \mathop{\inf}_{\phi,\psi_l \in [0,2\pi)} J(\phi, \valf, \vpsi, K), \numberthis \label{eq:optprob}
\end{flalign*}
where $J(\phi, \valf, \vpsi, K)$ is the objective function given by
\begin{align}
\label{eq:funobj}
J(\phi, \valf, \vpsi, K)= \!K\!\sum_{l=0}^{L-1} \abs{g_l(\phi)}^2 +  2 \sum_{l=0}^{L-2}\!\sum_{m=l+1}^{L-1}\!\abs{g_l(\phi)} \abs{g_m(\phi)}\!\mathop{\sum}_{k=0}^{K-1}\!\cos\!\lr{\psi_m-\psi_l - (\alpha_m-\alpha_l)kT}.
\end{align}

\begin{theorem}
\label{theorem:1}
The optimum of the objective function for an arbitrary $\phi$,
\begin{equation}
J^*(\phi)  \triangleq \sup_{\valf} \inf_{\vpsi} J(\phi, \valf, \vpsi, K), 
\end{equation}
is lower bounded as
\begin{equation}
\label{eq:genbound}    
J^*(\phi) \ge K\sum_{l=0}^{L-1} \abs{g_l(\phi)}^2 \qquad \text{when} \; L \le K,
\end{equation} 
and the solutions
\begin{align*}
\label{eq:optsol}
& \alpha^*_0=0, \; \lr{\lr{\alpha^*_m - \alpha^*_l}T/2} \in \mathcal{X}^*  \text{ for }  0\le l < m\le L-1, \\
&\qquad\qquad \mathcal{X}^* \triangleq\{q\pi/K: q\in\mathbb{Z}\} \setminus \{q \pi: q\in\mathbb{Z}\}, \numberthis
\end{align*}
guarantee the lower bound. A solution in~\eqref{eq:optsol} with the smallest possible nonnegative rates of phase shift is 
\begin{equation}
\label{eq:solLK}
\alpha^*_l = \frac{l2\pi}{KT}, \qquad l = 1, 2, \dots, L-1.
\end{equation} 
Furthermore, for $L=2$ and $3$, the bound in~\eqref{eq:genbound} is tight and the solutions in~\eqref{eq:optsol} are optimal.
\end{theorem}

\begin{IEEEproof}
See Appendix.
\end{IEEEproof}
\vspace{0.5\baselineskip}

\begin{example}
For $L=5$ and $K=5$, a solution set that achieves the lower bound in~\eqref{eq:J:star:bound} is given by $[\alpha^*_1,\alpha^*_2,\alpha^*_3,\alpha^*_4] =[2\pi/(KT),4\pi/(KT),6\pi/(KT),8\pi/(KT)]$ and $\alpha^*_0 = 0$ as the output of the $l=0$ antenna is not phase shifted.
\end{example}

In the case of $L>3$, proving the tightness of the bound in~\eqref{eq:J:star:bound} seems to be analytically intractable. In such a case, the optimization problem \eqref{eq:optprob} can be solved numerically when $L$ is not large.   

When the rates of phase shift $\alpha_l^*$ in~\eqref{eq:optsol} are used, the objective is independent of $\psi_l$ and hence independent of $\beta_l$.
Therefore, for any initial offset $\beta_l$ the worst-case \gls{AOA} $\phi$ that results in the highest \gls{BEP} is given by 
\begin{align}
\phi^{\star} = \mathop{\arg \min}_{\phi \in [0,2\pi)} \sum_{l=0}^{L-1} \abs{g_l(\phi)}^2.
\end{align}
As a consequence, when the proposed combining scheme is used to minimize the \gls{BEP} in case of multiple nonisotropic antennas, the antennas should be designed and oriented such that $\sum_{l=0}^{L-1} \abs{g_l(\phi^{\star})}^2$ is maximized.



\subsection{Two Antenna Case}
In this section, a few aspects specific to the $L=2$ antenna case are discussed.

\subsubsection{Different \gls{CAM} periods} 
\label{sec:diffCAM}        
the optimum rate of phase shift $\alpha^*=\alpha^*_1$ in the case of $L=2$ antennas for a given $K$ and $T$ has more than one solution given by \eqref{eq:optsol}. 
This allows a choice of $\alpha^*$ that is the optimum for several \gls{CAM} periods. 
Consider $R$ different periods where the $r$th period is given by $T_r = rT_1, \, \forall r \in \lrb{1,2,\dots,R}$.
The optimum rate of phase shift $\alpha^* = 2\pi/(KT_1)$ for the period $T_1$ is also the optimum for the other periods $T_r,\, \forall r \in \lrb{ \lrb{2,3,\dots,R} \setminus \lrb{K,2K,\dots} }$ since
\begin{align}
\alpha^* = \dfrac{2\pi}{KT_1} = \dfrac{2r\pi}{KT_r}.
\end{align}
The above $\alpha^*$ is not optimal for the periods $T_r, \forall r\in \lrb{K,2K,\dots}$ since \eqref{eq:optsol} is not satisfied.

\begin{example}
Let $T_1=0.1 \, \rm{s}$ and $T_2=0.3 \, \rm{s}$ be the periods of \glspl{CAM} arriving at the \gls{RX} from \gls{TX} 1 and \gls{TX} 2, respectively. 
Suppose $K=5$, the optimum rate of phase shift $\alpha^* = 2\pi/(KT_1)$ is optimum for both the periods.
\end{example}

\subsubsection{Similarity to \gls{EGC}}
in \gls{EGC}, signals from the two antennas are phase aligned or co-phased before they are added together to increase the \gls{SNR}.
This co-phasing can be achieved by phase shifting the output of the $l=1$ antenna $r_1(t)$ and adding it to the output of the reference antenna $r_0(t)$.
In the proposed combining scheme, $r_1(t)$ is phase shifted continuously and added to $r_0(t)$. 
When $\alpha^* = 2\pi/(KT)$, the signal corresponding to the $K$ consecutive packets at the $l=1$ antenna is shifted with $K$ different phases that uniformly sample the domain $[0,2\pi)$. 
Therefore, $\alpha^*$ minimizes the phase difference between the signals at the two antennas during one of the $K$ consecutive packets. 
The deviation from perfect co-phasing is dependent on the initial phase offset $\beta=\beta_1$.
As $K$ increases, the phase difference during one of the $K$ consecutive packets decreases and the output average \gls{SNR} of one of the $K$ packets reaches close to the case of \gls{EGC}.

\section{Comparison with Standard Schemes}
\label{sec:stdmethods}
In this section, the performance of the proposed combining scheme is compared with a few standard combining schemes.
The comparison is limited to the first scenario, where a single component arriving at an \gls{AOA} $\phi$ is considered.
The performance of the standard schemes in the second scenario when the channel at each antenna is independent complex \gls{AWGN} is well studied and can be found in~\cite[Sec. 7.2]{Goldsmith2005WL}.
In the case of the exponential \gls{PEP} function considered in Section~\ref{sec:burstprob}, minimizing the \gls{BEP} is equivalent to maximizing the sum of the average \glspl{SNR} of the $K$ packets as seen in~\eqref{eq:sumofsnrs}. 
Therefore, the sum of \glspl{SNR} is used as a performance criterion to compare the performance of the combining schemes.	 

\begin{enumerate}[leftmargin=*]

\item Single antenna: the sum of average \glspl{SNR} at the output of the $l$th antenna is given by
\begin{align*}
\rho_{l}(\phi) &= \mathop{\sum}_{k=0}^{K-1} \bar{\gamma}_{l}(\phi,k) = \frac{K P_{\rm{r}}}{P_{\rm{n}}} \abs{g_l(\phi)}^2.
\end{align*}
When the antenna is isotropic, the sum of average \glspl{SNR} is given by $\rho_{\rm{ISO}}(\phi)= K P_{\rm{r}}/ P_{\rm{n}}$.

\item \gls{MRC}: this scheme requires $L$ RF-chains and \glspl{ADC}, and a multiple port receiver that estimates the complex-valued channel gains and performs combining digitally.
The sum of average \glspl{SNR} is given by
\begin{align*}
\rho_{\rm{MRC}}(\phi)\!&=\!\mathop{\sum}_{k=0}^{K-1} \bar{\gamma}_{\rm{MRC}}(\phi,k)\!=\!\frac{K P_{\rm{r}}}{P_{\rm{n}}} \lr{\mathop{\sum}_{l=0}^{L-1}  \abs{g_l(\phi)}^2}.
\end{align*}

\item \gls{EGC}: this scheme requires $L$ RF-chains and \glspl{ADC}, and a multiple port receiver that estimates the channel phases and performs combining digitally.
The sum of average \glspl{SNR} is given by
\begin{align*}
\rho_{\rm{EGC}}(\phi)\!&=\!\mathop{\sum}_{k=0}^{K-1} \bar{\gamma}_{\rm{EGC}}(\phi,k)\!=\!\frac{K P_{\rm{r}}}{LP_{\rm{n}}} \lr{ \mathop{\sum}_{l=0}^{L-1} \abs{g_l(\phi)}}^2.
\end{align*}

\item \gls{SC}: this scheme requires $L$ RF-chains, and a digital or analog circuitry to measure the \glspl{SNR} on each branch and choose a branch.
The sum of average \glspl{SNR} is given by
\begin{align*}
\rho_{\rm{SC}}(\phi)\!&=\!\mathop{\sum}_{k=0}^{K-1} \bar{\gamma}_{\rm{SC}}(\phi,k)\!=\!\frac{K P_{\rm{r}}}{P_{\rm{n}}} \mathop{\max}_l \lrb{ \abs{g_l(\phi)}^2 }.
\end{align*}

\item \gls{ACN}: the proposed scheme requires analog phase shifters on $L-1$ branches operating independently and a combiner. 
The sum of average \glspl{SNR} when the solution in~\eqref{eq:optsol} is used is given by
\begin{align*}
\rho_{\rm{ACN}}(\phi) &= \mathop{\sum}_{k=0}^{K-1}\! \bar{\gamma}(\phi,\valf^*,\vbet ,k) \\
		   &= \frac{K P_{\rm{r}}}{LP_{\rm{n}}} \lr{ \mathop{\sum}_{l=0}^{L-1}{   \abs{g_l(\phi)}^2 }}, \forall \beta_l \in [0,2\pi).
\end{align*}
\end{enumerate}
The sum of average \glspl{SNR} in the case of \gls{MRC} and \gls{ACN} are relate as $\rho_{\rm{MRC}}(\phi) = L \rho_{\rm{ACN}}(\phi)$.

The \gls{MRC} scheme outperforms \gls{EGC}, \gls{SC}, and \gls{ACN} for any far-field functions $g_l(\phi)$. 
The relative performance of \gls{SC} and \gls{EGC} for an \gls{AOA} $\phi$ depends on the far-field functions $g_l(\phi)$.
The sum of average \glspl{SNR} of \gls{MRC}, \gls{EGC}, and \gls{SC} schemes is higher compared to our \gls{ACN} scheme, implying lower \gls{BEP}. 
However, these schemes require additional hardware and/or signal processing as mentioned above. 

\section{Numerical Results}

\begin{figure}
\centering
\includegraphics[scale=1]{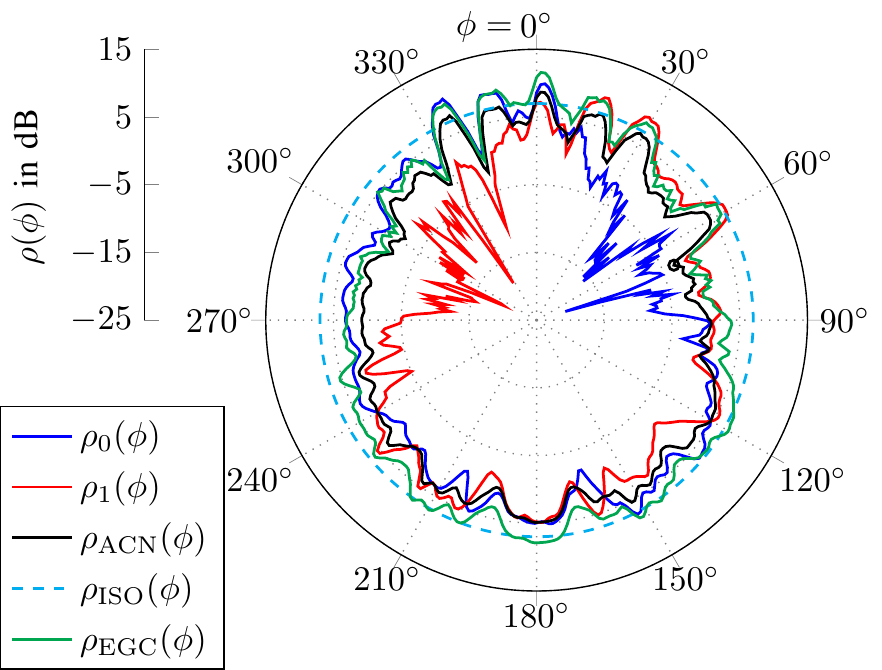}
\caption{$\rho(\phi)$ of the two monopoles mounted on the roof of a vehicle and the \gls{ACN}. The monopoles exhibit nonisotropic power gains. $P_{\rm{r}}/P_{\rm{n}}= 1$ and $K=5$.}
\label{fig:measured}
\end{figure}

\begin{figure}
\centering
\includegraphics[scale=1]{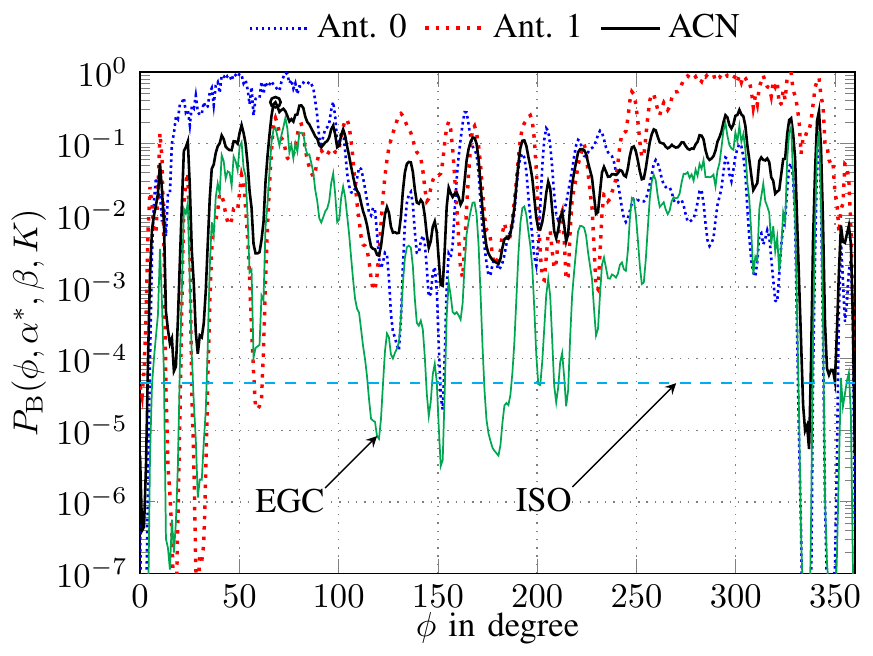}
\caption{\gls{BEP} as a function of \gls{AOA} $\phi$ for the individual antennas and the combined output when $K=5$, $\alpha^*=2\pi/(KT)$.}
\label{fig:burst}
\end{figure}

\begin{figure}
\centering
\includegraphics[scale=1]{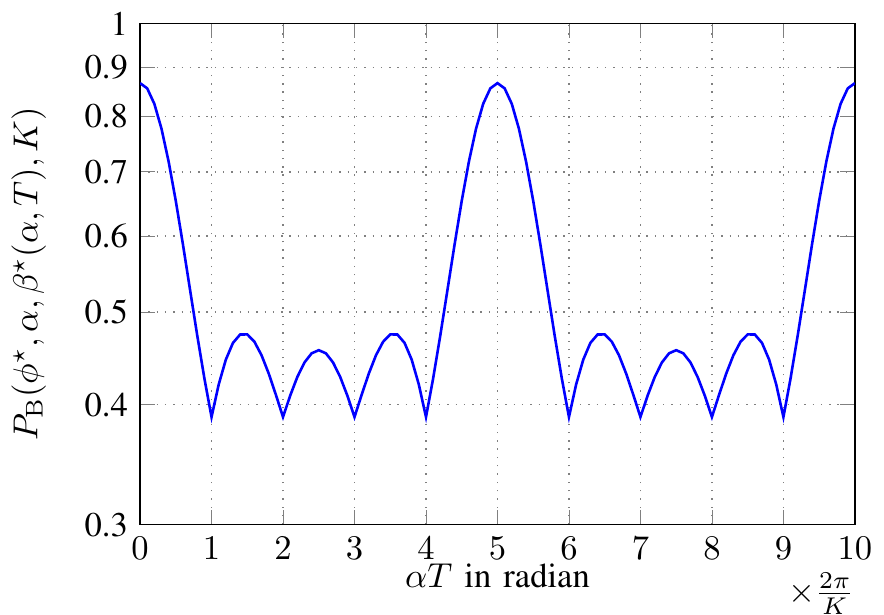}
\caption{\gls{BEP} as a function of $\alpha$ and $T$ for a fixed \gls{AOA} $\phi^{\star} \approx 68^{\circ}$ in the case of the two monopole antennas. $P_{\rm{r}}/P_{\rm{n}}= 1$ and $K=5$.}
\label{fig:missmatch}
\end{figure}

In this section, the performance of the \gls{ACN} is studied by using example antenna far-field functions.
The sum of average \glspl{SNR} $\rho(\phi)$ discussed in Section~\ref{sec:stdmethods} is used to illustrate the direction dependency of the \gls{BEP}. 
The $\rho_l(\phi)$ of the $l$th antenna is directly proportional to $\abs{g_l(\phi)}^2$ and therefore it also serves the purpose of visualizing the \gls{AOA} dependent gain of the antenna. 

The $\rho_0(\phi)$ and $\rho_1(\phi)$ of two monopole antennas placed on the roof of a Volvo XC90 are shown in \figref{fig:measured}.
The $l=0$ and $l=1$ monopole antennas are located at $(x,y)=(0,0.4 \,\rm{m})$ and $(0,-0.4 \,\rm{m})$, respectively (indicated by circles in \figref{fig:coord}). 
The $\rho(\phi)$ have been plotted by setting $P_{\rm{r}}/P_{\rm{n}} = 1$ and $K=5$.
The far-field function measurements were performed on the vertical polarization in the azimuth plane.
Therefore, we assume that the waves arriving in the azimuth plane have vertical polarization.
As seen in the figure, both the antennas exhibit very low $\rho(\phi)$ at certain \glspl{AOA}.
If only one of the two antennas is used, the packets arriving in the \glspl{AOA} of low $\rho(\phi)$ will have high \gls{BEP}.
The \gls{BEP} can be reduced by combining the output of the antennas using the proposed \gls{ACN}.
The performance of the \gls{ACN} is studied for $\alpha^* = \alpha^*_1 = 2\pi/(KT)$.
As seen in the figure, the $\rho_{\rm{ACN}}(\phi)$ of the \gls{ACN} has higher values for \glspl{AOA} where one of the two antennas has smaller values, implying lower \gls{BEP} at those \glspl{AOA}.
The sum of average \glspl{SNR} in the case of a single isotropic antenna and in the case of the measured antennas combined using \gls{EGC} are also shown in the figure for $K=5$.
The plots corresponding to \gls{MRC} and \gls{SC} have been omitted in the figure.
However, they are related to the plots in the figure through the relation $\rho_{\rm{MRC}}(\phi)=2 \rho_{\rm{ACN}}(\phi)$ and $\rho_{\rm{SC}}(\phi)= \max \lrb{\rho_{\rm{0}}(\phi),\rho_{\rm{1}}(\phi)}$.

\figref{fig:burst} shows the \gls{BEP} as a function of \gls{AOA} for the individual antennas and the \gls{ACN}.
The exponential \gls{PEP} function $P_{\mathrm{e}}(\bar{\gamma})=\exp(-\bar{\gamma}/5)$ is considered and $P_{\rm{r}}/P_{\rm{n}}= 10 \, \rm{dB}$ is used.
It is seen that the \gls{BEP} in the case of the individual antennas is very close to $1$ for the \glspl{AOA} that have very low $\rho_l(\phi)$ (see~\figref{fig:measured}).
The \gls{BEP} for the \glspl{AOA} corresponding to low gains in one of the two antennas is reduced by the \gls{ACN}. 
The \gls{BEP} for certain \glspl{AOA} when using the \gls{ACN} is higher in comparison to one of the individual antennas, this is expected as the \gls{ACN} operates without the knowledge of branch \glspl{SNR} and the complex-valued channel gains.
The figure also shows the \gls{BEP} in the case of a single isotropic antenna and in the case of the measured antennas combined using \gls{EGC}, the \gls{BEP} in these cases is in agreement with their $\rho(\phi)$ in \figref{fig:measured}.


\begin{figure}
\centering
\includegraphics[scale=1]{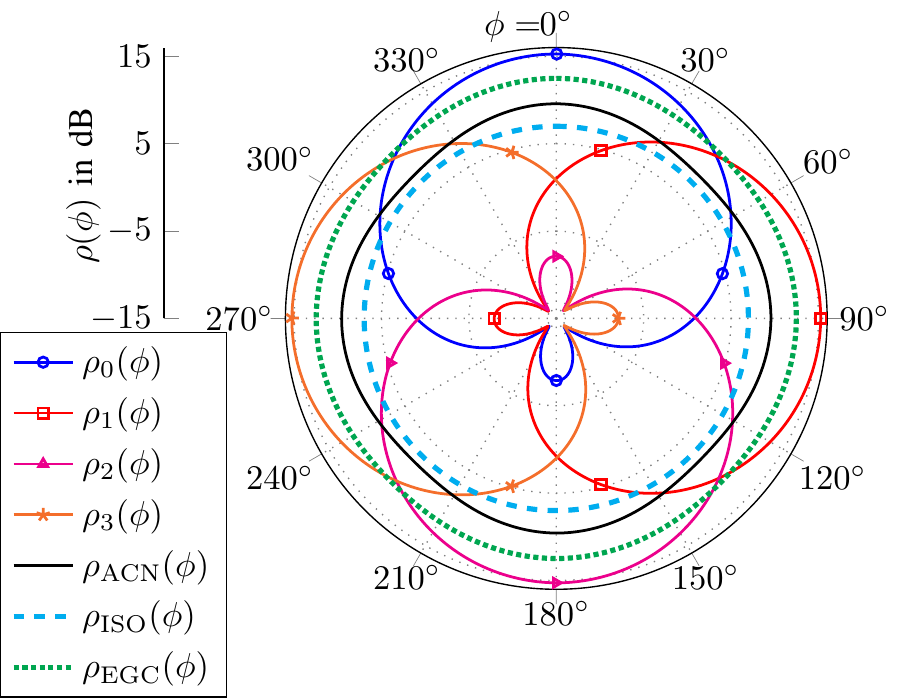}
\caption{$\rho(\phi)$ of the four patch antennas. The patch antennas exhibit nonisotropic power gains. $P_{\rm{r}}/P_{\rm{n}}= 1$ and $K=5$.}
\label{fig:patchrho}
\end{figure}

\begin{figure}
\centering
\includegraphics[scale=1]{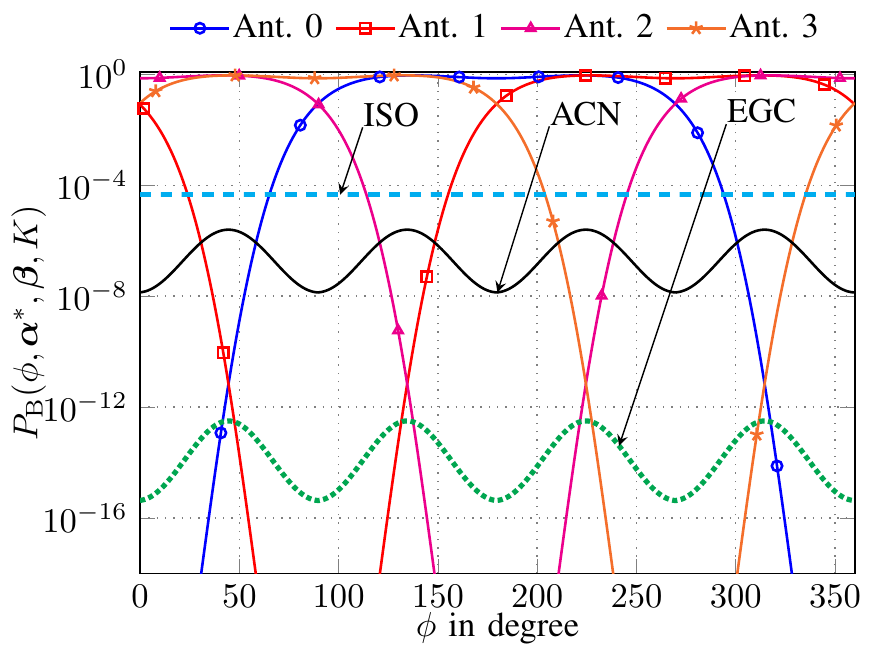}
\caption{\gls{BEP} as a function of \gls{AOA} $\phi$ for the individual antennas and the combined output when $K=5$, $[\alpha^*_1,\alpha^*_2,\alpha^*_3]=[2\pi/(KT),4\pi/(KT),6\pi/(KT)]$.}
\label{fig:patchburst}
\end{figure}

The performance of the \gls{ACN} when there is a mismatch in the period of \glspl{CAM} $T$ and/or the optimum rate of phase shift is shown in \figref{fig:missmatch} (note the multiplier on the horizontal axis).
The previous setup of the two monopole antennas with $K=5$ and $P_{\rm{r}}/P_{\rm{n}}= 10 \, \rm{dB}$ is used. 
The figure shows the \gls{BEP} as a function of $\alpha T$ for the worst-case \gls{AOA} $\phi^{\star} \approx 68^\circ$ (marked by a circle in \figref{fig:measured} and \figref{fig:burst}). 
The worst-case initial offset $\beta^{\star}(\alpha,T)$ is used for every $\alpha$ and $T$.
The \gls{BEP} is minimized when $\alpha T = (u2\pi)/K, \, u \in \lrb{1,2,\ldots} \setminus \lrb{K,2K,\ldots}$, which is in agreement with the solution in~\eqref{eq:optsol}. 
An $\alpha_0=2\pi/(KT_0)$ designed for $T=T_0$ is optimum for several integer multiples of $T_0$ and this result agrees with the discussion in Section~\ref{sec:diffCAM}. 
It can also be observed that the deviation of the \gls{BEP} from the minima is not significant for a large range of $\alpha T$.
Therefore, the \gls{ACN} can handle small mismatches in $\alpha$ or $T$ without significant performance loss.

As an example of $L>2$, we consider $L=4$ patch antennas.
The antennas $l=0,1,2$, and $3$ are located at $(x,y)=(1 \,\rm{m}, 0 \,\rm{m})$, $(0 \,\rm{m}, 0.6 \,\rm{m})$, $(-1 \,\rm{m}, 0 \,\rm{m})$, and $(0 \,\rm{m}, -0.6 \,\rm{m})$, respectively (indicated by squares in \figref{fig:coord}). 
The antennas are oriented such that the E-plane radiation pattern of each antenna coincides with the $xy$ plane and the perpendiculars to the ground planes pass through the origin of the coordinate system. 
The $\rho(\phi)$ of the antennas is shown in \figref{fig:patchrho}, $P_{\rm{r}}/P_{\rm{n}}= 1$ and $K=5$ are used.
The width, length, and height of the patch antennas are $0.5\lambda/\sqrt{\epsilon_{\rm{r}}}, \, 0.5\lambda/\sqrt{\epsilon_{\rm{r}}} $, and $0.05\lambda/\sqrt{\epsilon_{\rm{r}}}$, respectively,
where $\lambda$ is the wavelength of the carrier with frequency $5.9$ GHz and the dielectric constant of the substrate $\epsilon_{\rm{r}} = 2.2$.  
The length and width of the ground plane is equal to $\lambda$.
The far-field functions of the patch antennas are obtained using method of moments.
It can be observed that a single patch antenna exhibits very low $\rho(\phi)$ for a large range of \glspl{AOA} in the azimuth plane, implying higher \gls{BEP} at these \glspl{AOA}.
The \gls{ACN} can be used to combine the outputs of the four antennas to minimize the \gls{BEP} in these \glspl{AOA}.
The rates of phase shift for the three antennas are chosen according to \eqref{eq:solLK}, i.e.,  $\alpha^*_l = l2\pi/(KT), \, \text{for} \; l=1,2,3$.
It can be observed that $\rho_{\rm{ACN}}(\phi)$ of the \gls{ACN} has higher values for \glspl{AOA} where the individual antennas have very low values.
The sum of average \glspl{SNR} in the case of a single isotropic antenna and in the case of the patch antennas combined using \gls{EGC} are also shown in the figure for $K=5$.
The plots corresponding to \gls{MRC} and \gls{SC} have been omitted in the figure.
However, they are related to the plots in the figure through the relation $\rho_{\rm{MRC}}(\phi)=4\rho_{\rm{ACN}}(\phi)$ and $\rho_{\rm{SC}}(\phi)= \max \lrb{\rho_{\rm{0}}(\phi),\rho_{\rm{1}}(\phi),\rho_{\rm{2}}(\phi),\rho_{\rm{3}}(\phi)}$.
%

The \gls{BEP} in the setup of the four patch antennas as a function of \gls{AOA} is shown in \figref{fig:patchburst}.
An exponential \gls{PEP} function $P_{\mathrm{e}}(\bar{\gamma})=\exp(-\bar{\gamma}/5)$ is used and $P_{\rm{r}}/P_{\rm{n}}= 10 \, \rm{dB}$.
As in the case of $L=2$, the \gls{BEP} of the individual antennas is close to $1$ for the \glspl{AOA} that have very low $\rho(\phi)$. 
The \gls{BEP} for the \glspl{AOA} corresponding to low gains in the individual antennas is reduced by the \gls{ACN}. 
The \gls{ACN} enables robust communication for signals from all \glspl{AOA}.
The figure also shows the \gls{BEP} in the case of a single isotropic antenna and in the case of the patch antennas combined using \gls{EGC}, the \gls{BEP} in these cases is in agreement with their $\rho(\phi)$ in \figref{fig:patchrho}.

The optimization problem in~\eqref{eq:allpep} may not be analytically tractable for an arbitrary \gls{PEP} function, $K$, and $L$. 
In such a scenario, numerical optimization can be used to find the optimal rate of phase shifts for the \gls{ACN}.
We considered the optimization problem in the case of the two measured monopole antennas and $K=5$ with two \gls{PEP} functions for uncoded Gray-coded QPSK with independent bit errors~\cite[Ch. 6]{Goldsmith2005WL}, namely
\begin{align*}
\text{AWGN: } P_{\mathrm{e}}(\bar{\gamma}) &= 1-\lr{1 -  \qfunc{\sqrt{\bar{\gamma}}} }^{N_{\rm{b}}},\\
\text{Rayleigh fading: } P_{\mathrm{e}}(\bar{\gamma}) &= 1-\left( \frac{1}{2} + \frac{1}{2}  \sqrt{\frac{\bar{\gamma}}{2+\bar{\gamma}}}  \right)^{N_\mathrm{b}},
\end{align*}
where $N_\mathrm{b}$ is the number of bits in the packet and $\bar{\gamma}$ is the average \gls{SNR}.
Exhaustive search was used to solve the optimization problem numerically with $N_\mathrm{b}=3200$ and $P_{\rm{r}}/P_{\rm{n}}= 10 \, \rm{dB}$.
The analytically obtained optimum solution $\alpha^*= 2\pi/(KT)$ in the case of exponential \gls{PEP} function was found to be the optimum solution. 
We conjecture that the optimal solution in~\eqref{eq:optsol} is optimal for other monotonically decreasing \gls{PEP} functions.  

\section{Conclusion}
In this paper, we have proposed a simple method consisting of phase shifters to combine the outputs of $L$ nonisotropic antennas to enable robust vehicle-to-vehicle communications. 
To guarantee robustness, we have designed our method to minimize the burst error probability, i.e., the probability of $K$ consecutive packet errors for worst-case angle of arrivals.
The combining scheme does not need knowledge of the instantaneous complex-valued channel gains or the \glspl{SNR} on each antenna branch in contrast to the standard combining schemes. 
We have used measured radiation patterns of the antennas mounted on a vehicle and other example patterns to show the benefits of the scheme.

Rates of phase shift that guarantee an upper bound on the \gls{BEP} are derived in the case of $L \leq K$ and are given by $\alpha_l = l2\pi/(KT), \, \forall l=1,2,\ldots,L-1$.
Furthermore, it is shown that that the upper bound is indeed tight for the case of $L=2$ and $3$.  
Numerical results show that the proposed combining scheme can overcome the problem of a single nonisotropic antenna with very low gains in certain angle of arrivals.
In the case of two antennas, the optimum rate of phase shift designed for a specific $T$ is found to be robust to a large range of periods. 

The proposed scheme is also relevant for low cost sensor nodes with strict requirements on power consumption and complexity. Multiple low cost antennas with nonisotropic radiation patterns can be used and combined using the proposed method to support robust communications.

\appendix
We begin the proof of Theorem~\ref{theorem:1} by proving a few lemmas.
Define the function $f:\mathbb{R}^2\to\mathbb{R}$ as
\begin{equation}
  \label{eq:f:def}
  f(x, y) \triangleq \sum_{k=0}^{K-1} \cos(y - k 2 x),
\end{equation}
where $K > 1$ is a positive integer. It can be shown that
\begin{equation}
  \label{eq:f:simplified}
  f(x, y) =
  \begin{cases}
    K \cos(y), & x\in\mathcal{X}\\
    \displaystyle\frac{\sin(Kx)}{\sin(x)}\cos(y - (K-1)x), & x\notin\mathcal{X}
   \end{cases}
\end{equation}
where
\begin{equation}
  \label{eq:X:def}
  \mathcal{X} \triangleq\{q \pi: q\in\mathbb{Z}\}.
\end{equation}

\begin{lemma}
  \label{lemma:1}
  Let $f$ and $\mathcal{X}$ be as defined in~\eqref{eq:f:def} and~\eqref{eq:X:def}, respectively. Then,
  \begin{equation}
    \label{eq:f:zeros}
    f(x, y) = 0, \qquad x\in \mathcal{X}^*, y\in\mathbb{R},
  \end{equation}
  where
  \begin{equation}
    \label{eq:X:star:def}
    \mathcal{X}^* \triangleq\{q \pi/K: q\in\mathbb{Z}\}\setminus \mathcal{X}.
  \end{equation}
\end{lemma}
  \begin{IEEEproof}
    If $x\in\mathcal{X}^*$ then $x\notin\mathcal{X}$, and it follows from~\eqref{eq:f:simplified} that
    \begin{equation}
      f(x, y) = \frac{\sin(Kx)}{\sin(x)}\cos(y - (K-1)x),\qquad x\in\mathcal{X}^*,
    \end{equation}
    and since $\sin(Kx)/\sin(x) = 0$ for all $x\in\mathcal{X}^*$, the
    lemma follows.
  \end{IEEEproof}

\vspace{0.5\baselineskip}
\begin{lemma}
    \label{lemma:3}
    Let $\mathbf{x} \triangleq \lrh{x_1, x_2, \dots, x_{L-1}}^{\mathsf{T}}$, $\mathcal{X}^*$ be as defined in~\eqref{eq:X:star:def}, and let $x_0 = 0$.
    It is possible to find an $\mathbf{x}\in\mathbb{R}^{L-1}$ such that
    \begin{equation}
      \label{eq:main:condition}
      (x_m-x_l)\in\mathcal{X}^*, \qquad 0\le l < m\le L-1,
    \end{equation}
    if and only if
    $L\le K$. Moreover, one such construction is
    \begin{equation}
      \label{eq:x:construction}
      x_m = m\pi/K, \qquad m = 1, 2, \dots, L-1.      
    \end{equation}
\end{lemma}
  
  \begin{IEEEproof}
    We start by noting that, since $x_0 = 0$, the condition in~\eqref{eq:main:condition} is equivalent to the conditions
    \begin{align}
      \label{eq:equivalent:condition:1}
      (x_m-x_0) = x_m &\in\mathcal{X}^*, \qquad m = 1, 2, \dots, L-1\\
      \label{eq:equivalent:condition:2}
      (x_m-x_l)&\in\mathcal{X}^*, \qquad 1\le l < m \le L-1.
    \end{align}

    Now suppose $L\le K$ and let $x_m$ be as in~\eqref{eq:x:construction}.
    Since $1\le m \le L-1< K$, $m$ is not divisible by $K$. 
    Consequently, we have that $x_m = m\pi/K \in\mathcal{X}^*$ and ~\eqref{eq:equivalent:condition:1} is satisfied. 
    Moreover, for $1\le l < m \le L-1$,
    \begin{equation*}
      x_m-x_l\!=\!(m-l)\pi/K \in \{\pi/K, 2\pi/K, \dots, (L-2)\pi/K\},
    \end{equation*}
    which implies that~\eqref{eq:equivalent:condition:2} is satisfied. 
    Hence, we have shown that if $L\le K$, then there exists an $\mathbf{x}$ for which~\eqref{eq:main:condition} is satisfied.

    We will show that~\eqref{eq:main:condition}
    cannot be satisfied when $L>K$. We note
    that the condition in~\eqref{eq:equivalent:condition:1} is equivalent to
    $(x_m \in\mathcal{X}^*) \Leftrightarrow \lr{\mymod{x_m}{\pi} \in\mathcal{X}^{**}}$, where 
    \begin{equation}
      \label{eq:X:star:star:def}
      \mathcal{X}^{**}\triangleq\{\pi/K, 2\pi/K, \dots, (K-1)\pi/K\}
    \end{equation}
    and $\mymod{u}{v}$ is the remainder after dividing $u$ by $v$.
    Since the cardinality of $\mathcal{X}^{**}$ is $K-1$ and there are $L-1>K-1$ elements in $\mathbf{x}$, all which are members of $\mathcal{X}^{*}$, 
    there must exist a pair $(l, m)$ such that $ \mymod{x_m}{\pi} = \mymod{x_l}{\pi}$. 
    The existence of such a pair $(l, m)$ implies that $\mymod{[x_m - x_l]}{\pi} = 0\notin\mathcal{X}^{**}$,
    implying that $(x_m - x_l)\notin\mathcal{X}^*$, which violates the condition~\eqref{eq:equivalent:condition:2}.
    Hence, if $L>K$, it is not possible to find an $\mathbf{x}$ that satisfies~\eqref{eq:main:condition}. 
  \end{IEEEproof}

\vspace{0.5\baselineskip}

\begin{lemma}
\label{lemma:4}
   Let $f$ be as defined in~\eqref{eq:f:def}. If we can assign values to any $\ceil{W/2}$ of the elements in $\lrh{y_1,y_2,\dots,y_W}$, then we can satisfy the following condition
   \begin{equation}
   \label{eq:lemma4}
   	\sum_{w=1}^{W}  c_w f(x_w, y_w) \leq 0
   \end{equation}
   for an arbitrary $\mathbf{x}=\lrh{x_1,x_2,\dots,x_W}^{\mathsf{T}}$ and $c_w \in \mathbb{R}$ for $w=1,2,\dots,W$.
\end{lemma}

\begin{IEEEproof}
The sum in~\eqref{eq:lemma4} can be written as 
\begin{equation}    
\label{eq:lem4:sum}
    \sum_{w=1}^{W}  c_w f(x_w, y_w) = \sum_{w=1}^{W}  d_w \cos(y_w - e(x_w)),
\end{equation}
where
\begin{equation}    
    d_w(x_w) = \begin{cases}
    	  c_w K, & x\in\mathcal{X}  \\
    	  c_w (\sin(Kx_w)/\sin(x_w)), & x \notin \mathcal{X},	
          \end{cases}
\end{equation}
and 
\begin{equation}
e(x_w) = \begin{cases}
			0, & x\in\mathcal{X}  \\
			(K-1)x_w, & x \notin \mathcal{X}.	
		 \end{cases} 
\end{equation}

Define an one-to-one mapping $w \mapsto \tilde{w} \in \lrb{1,2,\dots,W}$ such that $\abs{d_{\tilde{w}=a}} \geq \abs{d_{\tilde{w}=b}}$ for $b>a$. The sum in~\eqref{eq:lem4:sum} can be split into two sums
\begin{align}
S_1 &= \sum_{\tilde{w}=1}^{\ceil{W/2}}     d_{\tilde{w}}(x_{\tilde{w}}) \cos(y_{\tilde{w}}-e(x_{\tilde{w}})) \\
S_2 &= \sum_{\tilde{w}= \ceil{W/2}+1}^{W}  d_{\tilde{w}}(x_{\tilde{w}}) \cos(y_{\tilde{w}}-e(x_{\tilde{w}})) 
\end{align}

If, for any $\mathbf{x} =\lrh{x_1, x_2, \dots, x_{W}}^{\mathsf{T}}$, $y_{\tilde{w}}$ can be chosen such that
\begin{equation}
\cos(y_{\tilde{w}}-e(x_{\tilde{w}})) = -\sgn(d_{\tilde{w}}) \, \text{for} \, \tilde{w} = 1,2,\dots,\ceil{W/2},
\end{equation}
then $S = S_1 + S_2 \leq 0$ and the lemma follows. 
\end{IEEEproof}
\vspace{\baselineskip}

The objective function in~\eqref{eq:funobj} can be written as
\begin{equation}
J(\phi, \valf, \vpsi, K) = K\sum_{l=0}^{L-1} \abs{g_l(\phi)}^2 
    + 2 \sum_{l=0}^{L-2} \sum_{m=l+1}^{L-1} |g_l(\phi)| |g_m(\phi)| f(x_m-x_l, y_m - y_l),
\end{equation}
where $x_l = \alpha_l T/2 \in \mathbb{R}$ and $y_l = \psi_l \in [0,2\pi)$.
Since $\alpha_0 = \beta_0 = 0$, we have that  $x_0 = 0$ and $y_0 = -\angle g_0(\phi)$.

For any $\phi$, the optimal value of the objective function,
\begin{align*}
J^*(\phi)  &\triangleq \sup_{\valf} \inf_{\vpsi} J(\phi, \valf, \vpsi, K) \\
&= K\sum_{l=0}^{L-1} \abs{g_l(\phi)}^2  + \mathop{\sup}_{\mathbf{x}} \mathop{\inf}_{\mathbf{y}} \sum_{l=0}^{L-2}\sum_{m=l+1}^{L-1} 2\abs{g_l(\phi)} \abs{g_m(\phi)} f(x_m-x_l, y_m - y_l), \numberthis \label{eq:optobj}
\end{align*}
where $\mathbf{x} \triangleq [x_1,x_2,\dots,x_{L-1}]^\mathsf{T}$ and $\mathbf{y} \triangleq [y_1,y_2,\dots,y_{L-1}]^\mathsf{T}$.

From Lemma~\ref{lemma:1}, we see that the second term in $J^*(\phi)$ is zero for any $\mathbf{y}$ if $(x_m-x_l)\in\mathcal{X}^*$,
for all pairs $(l, m)$ that occur in the double sum, i.e., for $0\le l < m\le L-1$.
It is shown in Lemma~\ref{lemma:3} that it is possible to find a solution that satisfies the aforementioned condition when $L \le K$. 
Therefore, we conclude that
\begin{equation}
    \label{eq:J:star:bound}
    J^*(\phi) \ge K\sum_{l=0}^{L-1} \abs{g_l(\phi)}^2, \qquad L \le K.
\end{equation} 

We now show that the bound in~\eqref{eq:J:star:bound} is tight for $L=2$ and $3$. The optimum objective in~\eqref{eq:optobj} can be written as
\begin{equation}
\label{eq:optmap}
J^*(\phi) = K\sum_{l=0}^{L-1} \abs{g_l(\phi)}^2 + \mathop{\sup}_{\mathbf{x}} \mathop{\inf}_{\mathbf{y}} \sum_{w=1}^{W} c_w f(\tilde{x}_w, \tilde{y}_w),
\end{equation}
where we have defined a mapping of the index pair $(l,m) \mapsto w \in \lrb{1,2,\dots,W}$ where $W=L(L-1)/2$ such that $c_w = 2\abs{g_l(\phi)} \abs{g_m(\phi)}$, $\tilde{x}_w=x_m-x_l$ and $\tilde{y}_w = y_m - y_l$.

As shown in Lemma~\ref{lemma:4}, for an arbitrary $\mathbf{x}$, if any $\ceil{W/2}$ of the elements in $[\tilde{y}_1,\tilde{y}_2,\dots,\tilde{y}_W]$ can be varied independently, then we can make
\begin{equation}
\label{eq:infneg}
\sum_{w=1}^{W} c_w f(\tilde{x}_w, \tilde{y}_w) \leq 0, \qquad\forall c_w \in \mathbb{R}.
\end{equation}

In the case of $L=2$, we have $\ceil{W/2}=1$ and $\tilde{y}_1 = y_1 - y_0$ can be varied independently by varying $y_1$. Therefore, the inequality in~\eqref{eq:infneg} holds.

In the case of $L=3$, the relation between $\tilde{y}_w$ and $\mathbf{y}$ in~\eqref{eq:optmap} is
\begin{equation}
\label{eq:independent}
 \begin{bmatrix}
          \tilde{y}_{0} \\
          \tilde{y}_{1} \\
          \tilde{y}_{2} 
        \end{bmatrix}
        = 
        \begin{bmatrix}
          1 & 0\\
          0 & 1\\
          -1 & 1
        \end{bmatrix}
         \begin{bmatrix}
          y_1 \\
          y_2 
        \end{bmatrix}
        -
        \begin{bmatrix}
          1\\
          1\\
          0
\end{bmatrix}
        y_0.
\end{equation}
It is easy to see that any $\ceil{W/2}=2$ rows in~\eqref{eq:independent} results in a consistent system of equations for solving for $\mathbf{y}$. Hence, any two of the three $\tilde{y}_w$ can be varied independently by varying $y_1$ and $y_2$. Therefore, the inequality in~\eqref{eq:infneg} holds.

Consequently, for $L=2$ and $3$, using \eqref{eq:infneg} in~\eqref{eq:optmap}, we have
\begin{equation}
\label{eq:ineqL3}
J^*(\phi) \leq K\sum_{l=0}^{L-1} \abs{g_l(\phi)}^2.
\end{equation}
Combining the results \eqref{eq:J:star:bound} and \eqref{eq:ineqL3} we conclude that
\begin{equation}
\label{eq:opt}
J^*(\phi) = K\sum_{l=0}^{L-1} \abs{g_l(\phi)}^2, \, L \leq K, \text{ and } L=2,3.
\end{equation}

This concludes the proof of Theorem~\ref{theorem:1}.

\bibliographystyle{IEEEtran}
\bibliography{Bibliography}
\end{document}